# 20 T DIPOLE MAGNET BASED ON HYBRID HTS/LTS COS-THETA COILS WITH STRESS MANAGEMENT*

A.V. Zlobin#, I. Novitski, E. Barzi, FNAL, Batavia, IL 60510, USA
P. Ferracin, LBNL, Berkeley, CA 94720, USA

## Abstract

This paper presents the design concept of the dipole magnet with 50 mm aperture, 20 T nominal field and 13% margin based on a six-layer cos-theta (CT) hybrid coil design. Due to the high stresses and strains in the coil at high field, Stress Management (SM) elements are implemented in the CT coil geometry. The results of magnet magnetic analysis are presented and discussed. The key parameters of this design are compared with the parameters of similar magnets based on block-type and canted cos-theta coils.

## INTRODUCTION

20 T dipole magnets are being considered for the next generation of particle accelerators. The 20 T field level is above the practical limit of $Nb_3Sn$ accelerator magnets and therefore it requires using superconducting materials with higher critical parameters such as High Temperature Superconductors (HTS). The high cost of HTS and the more complicated technology of HTS magnets make a hybrid approach, which uses both materials and technologies, attractive to minimize the volume of HTS coil. In a hybrid design, HTS materials are used in the high field part of the coil and $Nb_3Sn$ superconductors are used in the outer, lower-field coil regions. Several design options of 20 T dipole with a 50 mm clear aperture are being studied in the framework of US-MDP [1], including the Cos-theta (CT), Block-type (BL) and Common-Coil (CC) coil configurations [2], [3].

This paper presents a design concept of a dipole magnet with 50 mm aperture and 20 T nominal field with 13% margin based on the cos-theta (CT) coil design and a cold iron yoke. Due to the high stresses and strains in the coil at high fields, a Stress Management (SM) concept combined with the CT coil geometry is used. The results of magnet magnetic design optimization and analysis are presented and discussed. The key parameters of this design are compared with the parameters of similar magnet designs based on BL and CC coils.

## COIL DESIGN

The coil design optimization and magnetic calculations were performed using *ROXIE* [4] with a cylindrical cold iron yoke and real iron $B(H)$ curve. The presented magnet design is based on Rutherford cables made of Bi2212 (HTS) and $Nb_3Sn$ (LTS) composite superconducting wires. The strand and cable parameters are shown in Table 1. The cables parameters were selected based on the experience with multistrand Rutherford cables used in accelerator magnets and their further optimization during the magnet design process.

An optimized cross-section of the 20 T dipole coil is shown in Figure 1 and the coil geometrical parameters are summarized in Table 2. Coil turns in the coil layers are grouped into blocks separated by radial (interlayer) and azimuthal spacers used for the optimization of field quality in the aperture and for the mechanical stress management in the coil. The arrows in the coil blocks represent the relative value and direction of Lorents forces in each block.

Table 1. Strand and cable parameters

| Parameter | Cable 1 | Cable 2 | Cable 3 |
|---|---|---|---|
| Superconductor | Bi2212 | $Nb_3Sn$ | $Nb_3Sn$ |
| Strand diameter, mm | 1.0 | 1.0 | 0.7 |
| Cu/nonCu ratio | 3.0 | 1.1 | 1.1 |
| $J_c(15T;1.9K)$, A/mm² | 3750 | 2000 | 2000 |
| Number of strands | 32 | 40 | 40 |
| Cable width, mm | 16.5 | 20.1 | 15.0 |
| Cable small edge, mm | 1.85 | 1.70 | 1.22 |
| Cable large edge, mm | 1.95 | 1.90 | 1.38 |
| Cable packing factor | 0.83 | 0.90 | 0.81 |

Table 2. Coil parameters

| Parameter | Value |
|---|---|
| Number of layers | 6 |
| Number of blocks | 6 HTS+12 LTS |
| Number of turns/coil, L1-2/L3-4/L5-6 | 31/52/63 |
| Coil inner/outer diameter, mm | 50/310 |
| Bi2212 coil area/quadrant, mm² | 972 |
| $Nb_3Sn$ coil area/quadrant, mm² | 3110 |

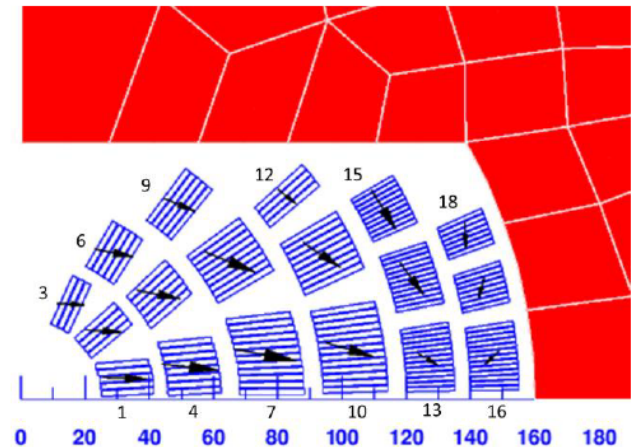

Figure 1: Cross-section of one quadrant of the 20 T dipole with cold yoke. Coil blocks are numbered and the Lorentz force vectors in the coil blocks are shown by the arrows.

___
* Work supported by Fermi Research Alliance, LLC, under contract No. DE-AC02-07CH11359 with the U.S. DOE and the U.S. Magnet Development Program (US-MDP).
# zlobin@fnal.gov

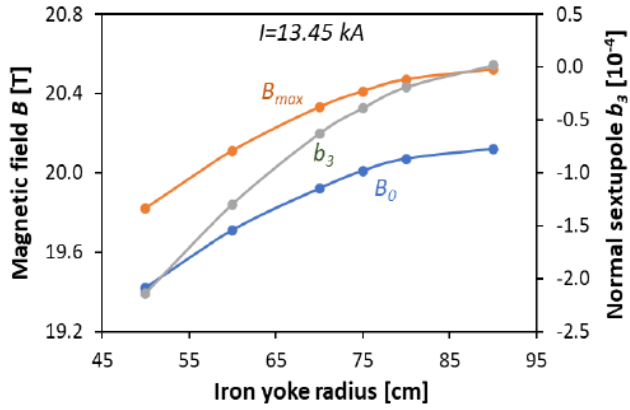

Figure 2: Effect of the iron yoke OD on $B_{max}$, $B_0$ and $b_3$ at a coil current of 13.45 kA.

Table 3. Geometrical field harmonics at $R_{ref}$=17 mm

| $n$ | 3 | 5 | 7 | 9 |
|---|---|---|---|---|
| $b_n$, $10^{-4}$ | -0.24 | 5.83 | 7.54 | -0.98 |

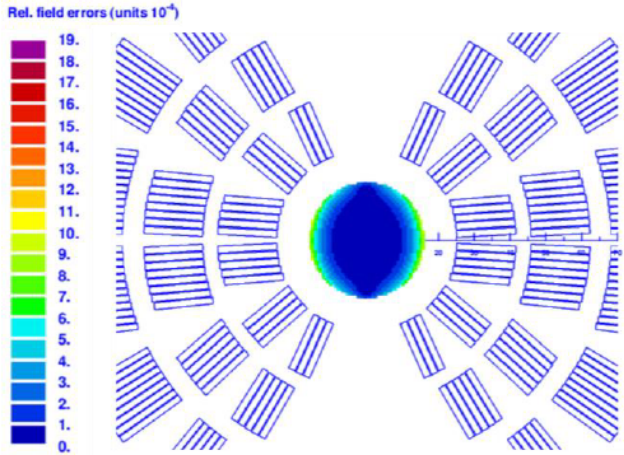

Figure 3: Field quality in the aperture within the 16 mm radius circle at a bore field of 20 T.

The effect of the iron yoke outer radius on the magnetic field value in the coil and in the aperture, as well as on the geometrical sextupole field component is shown in Figure 2. For yoke's outer radiuses above 75 cm the effect become smaller. The effect of the yoke inner shape was also estimated. With circular yoke inner surface, the magnet transfer function (TF) $B/I$ reduces by 3.3% with respect to the yoke inner shape shown in Figure 1. Based on these results, an iron outer radius of 75 cm and the iron inner shape shown in Figure 1 were selected for further analysis and optimization.

The low-order geometrical field harmonics in the magnet aperture at the reference radius $R_{ref}$ of 17 mm and a coil current of 13.45 kA, preliminary optimized for fields close to nominal, are summarized in Table 3. The field quality diagram in the magnet aperture at 13.45 kA current within a 16 mm radius circle is shown in Figure 3 (dark-blue area in the aperture). With still relatively large $b_5$ and $b_7$ the good quality field area, where $dB/B_1$ is smaller than 3 units, is close to 30 mm in diameter.

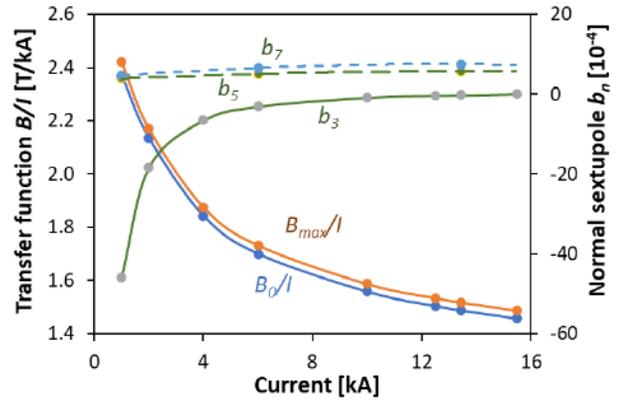

Figure 4: $B_{max}/I$, $B_0/I$ and $b_3$, $b_5$ and $b_7$ vs. coil current.

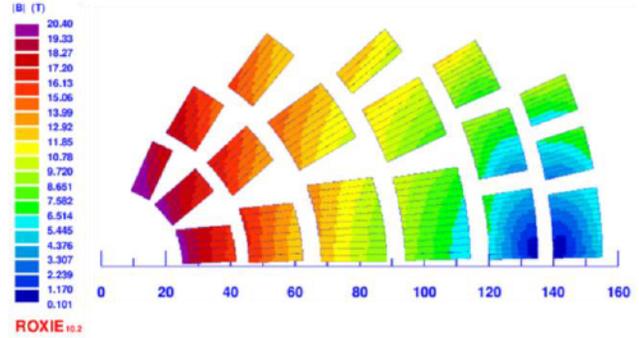

Figure 5: Magnetic field variation in the coil at $B_0$=20 T.

Table 4. Magnet parameters

| Parameter | Value |
|---|---|
| Coil nominal current $I_{nom}$, kA | 13.45 |
| Coil nominal field $B_{nom}$, T | 20.0 |
| Coil to aperture field ratio $B_{max}/B_0$ | 1.02 |
| Coil inductance @$I_{nom}$, mH/m | 52 |
| Stored energy @$I_{nom}$, MJ/m | 4.7 |
| Lorentz forces $F_x/F_y$ @$I_{nom}$, MN/m | 14.9/-7.4 |

Figure 4 shows variations of the magnet transfer functions (TF) $B_{max}/I$ and $B_0/I$ and the low-order field harmonics $b_3$, $b_5$ and $b_7$ vs the magnet current. Large variations of both TFs and $b_3$ with the current are due to the iron saturation effect. During the final magnet optimization this effect could be reduced by adding special holes in the yoke cross-section. The effect of the iron saturation on the higher-order harmonics $b_5$, $b_7$ and $b_9$ (not shown) is practically negligible. Minimization of these harmonics will be achieved by optimizing the number of turns in the coil blocks, and the block inclination angles and positions.

The optimized magnet parameters are summarized in Table 4. The magnet nominal design field of 20 T is achieved at a current of 13.45 kA. The magnet has a large stored energy of 4.7 MJ/m at the nominal field, and a relatively large inductance of 52 mH/m, which will require special analysis of magnet quench protection.

The field distribution in the coil at the nominal field of 20 T is shown in Figure 5. The maximum field for the three cables in the coil is reached in blocks 3 (Cable 1), 9 (Cable 2) and 15 (Cable 3) respectively.

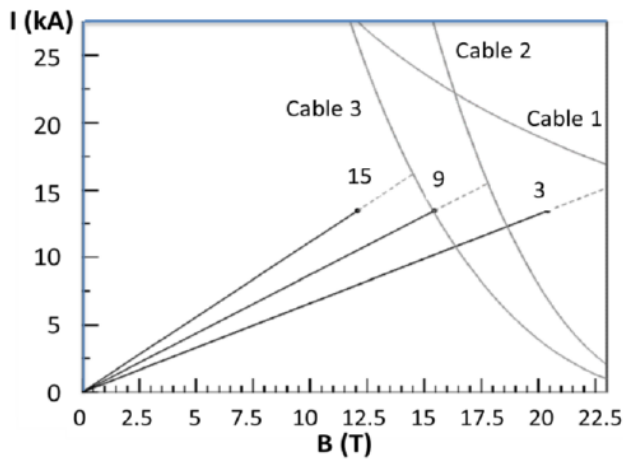

Figure 6: $I_c(B)$ curves at 1.9 K of the three cables used in the dipole, and load lines of the key coil blocks.

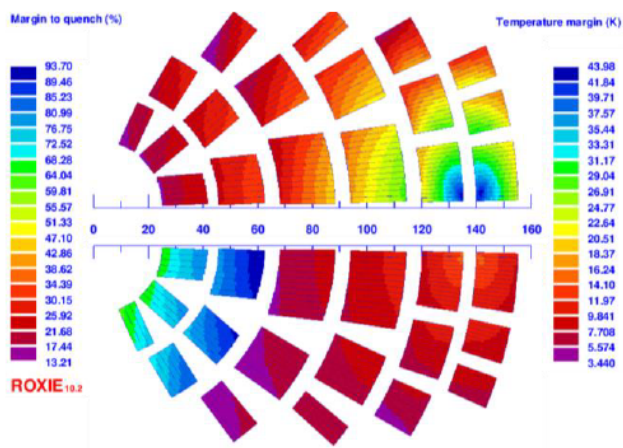

Figure 7: Margin to quench (top) and temperature margin (bottom) of the coil at the nominal field of 20 T and operation temperature 1.9 K.

Figure 6 shows the $I_c(B)$ dependences of the Bi2212 Cable 1 and Nb$_3$Sn Cables 2 and 3 at 1.9 K, and the maximum coil load lines for coil blocks 3, 9 and 15 representing the coils with different cable. The margin to quench for the three coils is 16.2% for Bi2212 coil (L1-2), 13.2% for Nb$_3$Sn Coil 2 (L3-4), and 16.9% for Nb$_3$Sn Coil 3 (L5-6).

Figure 7 shows the calculated margins to quench and temperature margin of different turns in the coil at the magnet operation temperature of 1.9 K and the nominal field of 20 T in the magnet aperture. The overall magnet quench margin determined by the Nb$_3$Sn Coil 2 (block 9) is 13.2%. This value meets the design criteria. Using advanced Nb$_3$Sn strands will allow increasing the total margin of this magnet design to 16%, which corresponds to a magnet quench field of 24 T. The minimum temperature margin for the Bi2212 coil is 28.65 K, and for the two Nb$_3$Sn coils it is 3.44 K and 4.23 K respectively.

Large Lorentz forces in this high-field dipole magnet lead to high mechanical stresses in the coil and to large coil block displacements with respect to their design positions. Since both Bi2212 and Nb$_3$Sn superconductors are brittle and may degrade or even lose their superconducting properties at stresses above 120 MPa and 150 MPa respectively, the stresses and displacements need to be kept within acceptable limits. This can be obtained by using special stress management elements and techniques under development for CT HTS and LTS coils [5]-[7]. Filling the radial and azimuthal spaces reserved between the coil blocks by a strong metallic structure will allow to control the coil mechanical stresses and block displacements in the coil in the whole range of magnet operation fields. These effects will be studied next in details using 2D and 3D mechanical analysis.

## CONCLUSION

A complementary conceptual design of a 20 T hybrid dipole demonstrator based on Bi2212 and Nb$_3$Sn coils has been developed and analyzed. The magnet provides a nominal target field of 20 T with 13.2% load line margins at 1.9 K with state-of-the-art superconductor parameters, six-layer hybrid shell-type coil and cold iron yoke.

The cross-section of the HTS coil and the total coil cross-section are noticeably reduced with respect to those in designs presented in [2], [3]. The decrease of coil volume and especially of the HTS coil cross-section will allow reducing the cost of this 20 T hybrid dipole.

The stress management elements are being integrated to the coil cross-section to keep the mechanical stresses in brittle Bi2212 and Nb$_3$Sn superconductors below their dangerous level. The optimization of the stress management elements in this magnet design needs to be coordinated with the practical results of magnet R&D programs.

Further magnet design optimization, including field quality, operation margins, stress level will be done in the next design study phase.